\documentclass[prc,aps,floatfix,groupedaddress,amsmath,amssymb,twocolumn]{revtex4}
\usepackage{graphicx}
\usepackage{epsfig}
\usepackage{dcolumn}
\usepackage{bm}
\usepackage{color}
\usepackage[normalem]{ulem}

\begin{document}

\title{Charmed nuclei within a microscopic many-body approach}
\author{I. Vida\~na$^1$, A. Ramos$^2$ and C. E. Jim\'enez-Tejero$^3$}  
\affiliation{$^1$Istituto Nazionale di Fisica Nucleare, Sezione di Catania, Dipartimento di Fisica ``Ettore Majorana", Universit\`a di Catania, Via Santa Sofia 64, I-95123 Catania, Italia}
\affiliation{$^2$Departament de F\'{i}sica Qu\`antica i Astrof\'{i}sica and Institut de Ci\`encies del Cosmos (ICCUB), Facultat de F\'{i}sica, Universitat de Barcelona, Mart\'{i} i Franqu\`es 1, E-08028 Barcelona, Spain}
\affiliation{$^3$Barcelona Center for Subsurface Imaging, Institute of Marine Sciences, Spanish Research Council, Passeig Mar\'{i}tim de la Barceloneta 37, E-08003 Barcelona, Spain}
\newcommand{\m}{\multicolumn}
\renewcommand{\arraystretch}{1.2}

\begin{abstract}

Single-particle energies of the $\Lambda_c$ chamed baryon are obtained in several nuclei from the relevant self-energy constructed
within the framework of a perturbative many-body approach. Results are presented for a charmed baryon-nucleon ($Y_cN$) potential based on a SU(4)
extension of the meson-exchange hyperon-nucleon potential $\tilde A$ of the J\"{u}lich group. Three different models (A, B and C) of this interaction, that differ only on the values of the couplings of the scalar $\sigma$ meson with the charmed baryons, are considered. Phase shifts, scattering lengths and effective ranges are
computed for the three models and compared with those predicted by the $Y_cN$ interaction derived in Eur. Phys. A {\bf 54}, 199 (2018) from the extrapolation to the physical pion mass of recent results of the HAL QCD Collaboration. Qualitative agreement is found for two of the models (B and C) considered. Our results for $\Lambda_c$-nuclei are compatible with those obtained by other authors based on different models and methods. We find a small spin-orbit splitting of the $p-, d-$ and $f-$wave states as in the case of single $\Lambda$-hypernuclei. The level spacing of $\Lambda_c$ single-particle energies is found to be smaller than that of the corresponding one for hypernuclei. The role of the Coulomb potential and the effect of the coupling of the $\Lambda_cN$ and $\Sigma_cN$ channels on the single-particle properties of $\Lambda_c-$nuclei are also analyzed. Our results show that, despite the Coulomb repulsion between the $\Lambda_c$ and the protons, even the less attractive one of our $Y_cN$ models (model C) is able to bind the $\Lambda_c$ in all the nuclei considered. The effect of the $\Lambda_cN-\Sigma_cN$ coupling is found to be almost negligible due to the large mass difference of the $\Lambda_c$ and $\Sigma_c$ baryons.

\end{abstract}

\maketitle

\section{Introduction}

Soon after the discovery of charmed hadrons \cite{aubert74,augustin74,cazzoli75,goldhaber76,peruzzi76,knapp76}, the possible existence of charmed nuclei (bound 
systems composed of nucleons and charmed baryons) was proposed in analogy 
with hypernuclei (see {\it e.g.,} Refs.\ \cite{tyapkin75,dover77a,dover77b,iwao77,gatto78}). This possibility motivated several authors to study the properties of these systems within different theoretical 
approaches, predicting a rich spectrum and a wide range of atomic numbers \cite{bhamathi81,kolesnikov81,bando82a,bando82b,bando82c,gibson83,bhamathi89}. 
Production mechanisms of charmed nuclei  by means of {\it charm exchange} or {\it associate charm production} reactions, analogous 
to the ones widely used in hypernuclear physics, were also proposed \cite{bressani89,bunyatov91}. However, the experimental production 
of charmed nuclei is difficult and, up to now, only three ambiguous candidates have been reported by an emulsion experiment carried 
out in Dubna in the mid-1970s \cite{batusov81a,batusov81b,batusov81c,batusov81d,lyukov89}. Experimental difficulties arise mainly from (i) the kinematics of the production 
reactions: charmed particles are formed with large momentum making their capture by a target-nucleus improbable; and (ii) the short 
lifetimes of $D$-meson beams, which makes necessary to place the target as close as possible to the $D$-meson production point. Such 
difficulties will be hopefully overcome at the future GSI--FAIR (Gesellschaft f\"{u}r Schwerionenforschung--Facility for Antiproton 
and Ion Research) and JPARC (Japan Proton Accelerator Research Complex) facilities \cite{riedl07,Tomasi-Gustafsson:2018fwm}. The production of charmed particles 
in these facilities will be sufficiently large to make the study of charmed nuclei possible . 
Studies of ${\bar p}$ reactions in nuclei at the conditions of the $\overline {\rm P}$ANDA experiment predict forward differential cross sections for the formation of $\Lambda_c$ hypernuclei in the range of a few $\mu$b/sr \cite{Shyam:2016uxa}.
These future prospects  
have injected a renewed interest in this line of research \cite{Krein:2017usp}. In the last few years, theoretical estimations of the charmed baryons 
properties in nuclear matter and finite nuclei have been revisited using the quark-meson coupling model \cite{tsushima03a,tsushima03b,tsushima03c,tsushima03d}, 
a relativistic mean field approach \cite{tan04}, effective Lagrangians satisfying the heavy quark, chiral and hidden local 
symmetries \cite{liu11}, the quark cluster model \cite{Maeda:2015hxa}, or a single-folding potential employing a Lattice QCD (LQCD) simulation of the $\Lambda_c N$ interaction  \cite{Miyamoto:2017tjs}. An extrapolation to the physical pion mass of the former LQCD $\Lambda_c N$ interaction has recently become available \cite{Haidenbauer:2017dua}.

In this work we study the single-particle properties of the $\Lambda_c$ charmed baryon in several nuclei using 
a microscopic many-body approach. Our starting point is a nuclear matter $G$-matrix derived from a bare charmed 
baryon-nucleon ($Y_cN, Y_c=\Lambda_c,\Sigma_c$) potential based on a SU(4) extension of the hyperon-nucleon ($YN$) potential 
$\tilde A$ of the J\"{u}lich group \cite{reuber94}. This $G$-matrix is used to calculate the 
self-energy of the $\Lambda_c$ in the finite nucleus including corrections up to the second order. Solving the Schr\"{o}dinger 
equation with this self-energy 
we are able to determine the single-particle energies and the wave function of the bound $\Lambda_c$. Our approach
also provides the real and imaginary parts of the $\Lambda_c$ optical potential at positive energies, and 
therefore, allows one to study the $\Lambda_c$-nucleus scattering properties. This method was already used to study 
the properties in finite nuclei of the nucleon \cite{borromeo92}, the $\Delta$ isobar \cite{morten94}, and the $\Lambda$ and $\Sigma$ 
hyperons \cite{morten96,vidana98,vidana00,vidana17}.

The paper is organized in the following way. In Sec.\ \ref{sec:sec2} we present our model for the $Y_cN$ interaction. The method 
to obtain the $\Lambda_c$ single-particle properties in finite nuclei is briefly described in Sec.\ \ref{sec:sec3}. Results for 
a variety of $\Lambda_c-$nuclei are shown in Sec.\ \ref{sec:sec4}. Finally, a brief summary and some concluding remarks 
are given in Sec.\ \ref{sec:sec5}.

\section{The $Y_cN $ interaction}
\label{sec:sec2}

\begin{figure}[t]
\begin{center}
\includegraphics[width=5.0cm]{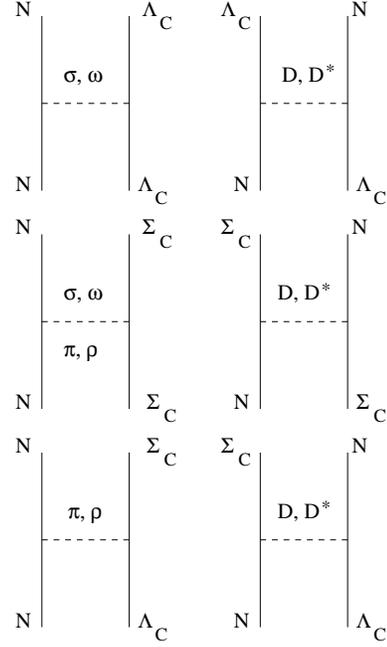}
\caption{Single-meson exchange contributions included in our model for the $Y_cN$ interaction.}
\label{fig:ycnpot}
\end{center}
\end{figure}

Our model for the $Y_cN$ interaction is based on a generalization of the meson exchange $YN$
potential $\tilde A$ of the J\"{u}lich group \cite{reuber94}. In analogy with that model, we describe the three different channels,
 $\Lambda_cN \rightarrow\Lambda_cN$, $\Sigma_cN \rightarrow\Sigma_cN$ and $\Lambda_cN \leftrightarrow\Sigma_cN$,
only as the sum of single scalar, pseudoscalar and vector meson exchange potentials shown in Fig.\ \ref{fig:ycnpot}. As in the $YN$ J\"{u}lich 
potential, the exchange of the effective scalar $\sigma$ meson parametrizes the contribution of  correlated $2\pi$-exchange.
The basic input of our model are the baryon-baryon-pseudoscalar (BBP) and the baryon-baryon-vector (BBV) vertices 
described, respectively, by the Lagrangian densities
\begin{eqnarray}
{\cal L}_{BBP}&=& g_{NN\pi} (N^\dag \vec \tau N) \cdot  \vec \pi \nonumber \\
             &+& g_{\Lambda_c \Sigma_c \pi}[\vec \Sigma_c^\dag \cdot \vec \pi  \Lambda_c+\Lambda_c^\dag\vec \Sigma_c \cdot \vec \pi] \nonumber \\
             &-i&g_{\Sigma_c\Sigma_c\pi} (\vec \Sigma_c^\dag \times \vec \Sigma_c) \cdot \vec \pi \nonumber \\
             &+& g_{N\Lambda_c D}[(N^\dag D)\Lambda_c +\Lambda_c^\dag( D^\dag N)] \nonumber \\
             &+& g_{N\Sigma_c D}[(N^\dag \vec \tau D)\cdot \vec \Sigma_c +\vec \Sigma_c^\dag( D^\dag \vec \tau N)] \nonumber \
\label{eq:BBP}
\end{eqnarray}
and 
\begin{eqnarray}
{\cal L}_{BBV}&=& g_{NN\rho} (N^\dag \vec \tau  N) \cdot  \vec \rho \nonumber \\
             &+& g_{\Lambda_c \Sigma_c \rho}[\vec \Sigma_c^\dag \cdot \vec \rho  \Lambda_c+\Lambda_c^\dag\vec \Sigma_c \cdot \vec \rho\,] \nonumber \\
             &-i&g_{\Sigma_c\Sigma_c\rho} (\vec \Sigma_c^\dag \times \vec \Sigma_c) \cdot \vec \rho \nonumber \\
             &+& g_{N\Lambda_c D^*}[(N^\dag D^*)\Lambda_c +\Lambda_c^\dag( D^{*\dag} N)] \nonumber \\
             &+& g_{N\Sigma_c D^*}[(N^\dag \vec \tau D^*)\cdot \vec \Sigma_c +\vec \Sigma_c^\dag( D^{*\dag} \vec \tau N)] \nonumber \\
             &+& g_{NN\omega}N^\dag N \omega \nonumber \\
             &+& g_{\Lambda_c\Lambda_c\omega} \Lambda_c^\dag\Lambda_c \omega\nonumber \\
             &+& g_{\Sigma_c\Sigma_c\omega} \vec \Sigma_c^\dag \cdot \vec \Sigma_c \omega \nonumber \ .
\label{eq:BBV}
\end{eqnarray}
We note that the isospin structure of these vertices is the same as that of their analogous strange ones. Similarly to the J\"{u}lich $YN$ interaction, which is 
itself based on the Bonn NN one, the $Y_c N$ model presented here also neglects the contribution of the $\eta$ and $\eta'$ 
mesons. 

We use the SU(4) symmetry to derive the relations between the different coupling constants. Note, however, that this symmetry
is strongly broken due to the use of the different physical masses of the various baryons and mesons, and that 
we employ it rather as a mathematical tool to get a handle on the various couplings of our model. In particular, we are dealing with 
$J^P=\frac{1}{2}^+$ baryons and $J^P=0^-,1^-$ mesons belonging to  $20'$- and $15$-plet irreducible representations 
of SU(4), respectively. Since the baryon current can be reduced according to
\begin{equation}
{\bf 20'} \otimes {\bf \overline{20'}} = {\bf 1} \oplus {\bf 15_1} \oplus {\bf 15_2} \oplus {\bf 20''} 
\oplus {\bf 45} \oplus {\bf \overline{45}} \oplus {\bf 84} \oplus {\bf 175} \ ,
\label{eq:20x20}
\end{equation}
there are two ways to obtain an SU(4)-scalar for the coupling ${\bf 20'} \otimes {\bf \overline{20'}} \otimes {\bf 15}$ because 
the baryon current contains two distinct $15$-plet representations, ${\bf 15_1}$ and ${\bf 15_2}$. They couple
to the meson $15$-plets with strengh $g_{15_1}$ and $g_{15_2}$, respectively. It is quite straightforward to relate 
these two couplings to the couplings $g_D$ and $g_F$ of the usual symmetric 
(``D-coupling") and antisymmetric (``F-coupling") octet representations of the baryon current in SU(3). They read
\begin{eqnarray}
g_{15_1}&=&\frac{1}{4}(7g_D+\sqrt{5}g_F)=\sqrt{\frac{10}{3}}g_8(7-4\alpha) \nonumber \\
g_{15_2}&=&\sqrt{\frac{3}{20}}(\sqrt{5}g_D-5g_F)=\sqrt{40}g_8(1-4\alpha) \ ,
\label{eq:g15}
\end{eqnarray}
where in the last step we have written $g_D$ and $g_F$ in terms of the conventional SU(3) octet strengh coupling $g_8$, and the so-called 
$F/(F+D)$ ratio $\alpha$
\begin{equation}
g_D=\frac{40}{\sqrt{30}}g_8(1-\alpha) \ , \,\,\,\,\,\, g_F=4\sqrt{6}g_8\alpha \ .
\label{eq:g8}
\end{equation}

\begin{table}
\begin{center}
\begin{tabular}{ccccc}
\hline
\hline
Model &Vertex & $g_{BBM}/\sqrt{4\pi}$ & $f_{BBM}/\sqrt{4\pi}$ & $\Lambda_{BBM}$ (GeV) \\ 
\hline
A,B,C &$NN\pi$                              &  $3.795$   & $-$ & $1.3$ \\ 
A,B,C &$\Lambda_c\Sigma_c\pi$   &  $3.067$   & $-$ & $1.4$ \\ 
A,B,C &$\Sigma_c\Sigma_c\pi$      &  $2.277$   & $-$ & $1.2$ \\  \\

A,B,C &$N\Lambda_c D$                & $-3.506$ & $-$ & $2.5$ \\ 
A,B,C &$N\Sigma_c D$                   & $1.518$   & $-$ & $2.5$ \\  \\

A,B,C &$NN\rho$                & $0.917$ & $5.591$ & $1.4$ \\ 
A,B,C &$\Lambda_c\Sigma_c\rho$ & $0.000$ & $4.509$ & $1.16$ \\ 
A,B,C &$\Sigma_c\Sigma_c\rho$  & $1.834$ & $3.372$ & $1.41$ \\ \\ 

A,B,C &$NN\omega$                 & $4.472$  & $0.000$  & $1.5$ \\ 
A,B,C &$\Lambda_c\Lambda_c\omega$ & $1.490$  & $2.758$  & $2.0$ \\ 
A,B,C &$\Sigma_c\Sigma_c\omega$   & $1.490$  & $-2.907$ & $2.0$ \\ \\

A,B,C &$N\Lambda_c D^*$        & $-1.588$ & $-5.175$ & $2.5$ \\ 
A,B,C &$N\Sigma_c D^*$         & $-0.917$ & $2.219$ & $2.5$ \\ \\

A,B,C&$NN\sigma$                       & $2.385$         &  $-$ & $1.7$ \\ \\

A&$\Lambda_c\Lambda_c\sigma$       & $2.138$  & $-$  & $1.0$ \\
A&$\Sigma_c\Sigma_c\sigma (I=1/2)$ & $3.061$  & $-$ & $1.0$ \\
A&$\Sigma_c\Sigma_c\sigma (I=3/2)$ & $3.102$  & $-$  & $1.12$\\ \\

B&$\Lambda_c\Lambda_c\sigma$       & $1.817$  & $-$  & $1.0$ \\
B&$\Sigma_c\Sigma_c\sigma (I=1/2)$ & $2.601$  &  $-$ & $1.0$ \\
B&$\Sigma_c\Sigma_c\sigma (I=3/2)$ & $2.636$  &  $-$ & $1.12$\\ \\

C&$\Lambda_c\Lambda_c\sigma$       & $1.710$  & $-$  & $1.0$ \\
C&$\Sigma_c\Sigma_c\sigma (I=1/2)$ & $2.448$  & $-$  & $1.0$ \\
C&$\Sigma_c\Sigma_c\sigma (I=3/2)$ & $2.481$  & $-$  & $1.12$\\ 

\hline
\hline
\end{tabular}
\end{center}
\caption{Baryon-baryon-meson coupling constants $g_{BBM}$, $f_{BBM}$ and cutoff masses $\Lambda_{BBM}$ for the models A, B and C of the $Y_cN$ interaction constructed and used in this work.}
\label{tab:tab1}
\end{table}


Let us consider first the coupling of the baryon current to the pseudoscalar mesons. The relations between 
all the relevant BBP coupling constants can be easily obtained by using SU(4) Clebsch--Gordan cofficients 
\cite{haacke76} and the above relations. They read
\begin{eqnarray}
&g_{\Lambda_c\Sigma_c\pi}&=\frac{2}{\sqrt{3}}\,g_{NN\pi}(1-\alpha_p) \nonumber \\
&g_{\Sigma_c\Sigma_c\pi}&=2\,g_{NN\pi}\alpha_p \nonumber \\
&g_{N\Lambda_c D}&=-\frac{1}{\sqrt{3}}g_{NN\pi}(1+2\alpha_p) \nonumber \\
&g_{N\Sigma_c D}&=g_{NN\pi}(1-2\alpha_p) \ ,
\label{eq:su4cc1}
\end{eqnarray}
where we have added the subindex $p$ to the ratio $\alpha$ to specify that this is the ratio for the coupling of baryons with the pseudoscalar mesons and distinguish it from that for the vector ones used below.

Similarly, the corresponding relations for the BBV couplings can be obtained by
simply making the replacements $\pi\rightarrow\rho, D\rightarrow D^*, \alpha_p\rightarrow \alpha_v$
in the above expressions. In addition, the couplings to the $\omega$ meson are
\begin{eqnarray}
&g_{NN\omega}&=g_{NN\rho}(4\alpha_v-1) \nonumber \\
&g_{\Lambda_c\Lambda_c\omega}&= \frac{g_{NN\rho}}{9}(6\alpha_v+3)  \nonumber \\
&g_{\Sigma_c\Sigma_c\omega}&= g_{NN\rho}(2\alpha_v-1) \ , 
\label{eq:su4cc2}
\end{eqnarray}
where we have assumed that the physical $\omega$ meson results from the ideal mixing of the mathematical
members of the 15-plet $\omega_8$ and $\omega_1$. 

The relations for the tensor coupling constants $f_{BBM}$ can be obtained by applying the corresponding 
SU(4) relations to the ``magnetic" coupling $G_{BBM}=g_{BBM}+f_{BBM}$. Thus, in the above relations $g_{v}$ has
to be replaced simply by $G_{v}$ and $\alpha_v$ by $\alpha_t$.  

To determine the couplings of the scalar $\sigma$ meson with the charmed baryons, we should remind that this meson is not a member of any SU(4) multiplet and, therefore, it is not possible to obtain these couplings by invoking the SU(4) symmetry as we did for the couplings with the pseudoscalar and vector mesons. This leaves us certain freedom to chose the values of the couplings $g_{\Lambda_c\Lambda_c\sigma}$ and 
$g_{\Sigma_c\Sigma_c\sigma}$. To explore the sensitivity of our results to these couplings, in this work we consider three different sets of values for them that, together with those for the pseudoscalar and vector meson couplings, define three models for the $Y_cN$ interaction. From now on we will refer to these models simply as A, B and C. In model A the couplings of the $\sigma$ meson with the charmed baryons are assumed to be equal to its couplings with the $\Lambda$ and $\Sigma$ hyperons, and their values are taken from the original $YN$ potential $\tilde A$ of the J\"{u}lich group \cite{reuber94}. In models B and C these couplings are reduced by 15\% and 20\%, respectively, with respect to model A. The coupling $g_{NN\sigma}$  have been taken, for the three models, equal to that of the J\"{u}lich $\tilde A$ $YN$ potential.
 
Taking the values $\alpha_P=0.4$, $\alpha_v=1$ and $\alpha_t=0.4$  employed in \cite{reuber94}, we obtain the couplings reported in Table~\ref{tab:tab1} where we also show the cutoff masses $\Lambda_{BBM}$  of the monopole form factors of the different vertices. We note that, to describe the nucleon-nucleon data quantitatively, the coupling $g_{NN\omega}$ in the J\"{u}lich $\tilde A$ $YN$ model was increased by a factor 1.626 with respect to its SU(3) value, $g_{NN\omega}=3 g_{NN\rho}$, thereby accounting for missing short-range correlations in an effective way. In the present work, we apply the same increasing factor to the $g_{\Lambda_c\Lambda_c\omega}$ and $g_{\Sigma_c\Sigma_c\omega}$ coupling constants of Eq.~(\ref{eq:su4cc2}). We note also that the relation of these coupling constants to  $g_{NN\rho}$ is a factor of two smaller than that obtained in the SU(3) sector, while the relations in Eq.~(\ref{eq:su4cc1}), involving charmed baryons and the $\pi$, $\rho$, $D$ and $D^*$ mesons, are the same as those involving their counterparts in the strange sector. 

\begin{figure}[t]
\begin{center}
\includegraphics[width=9.0cm]{fig2.eps}
\caption{(color on-line) $^1$S$_0$ and $^3$S$_1$  $\Lambda_cN$ phase shift as a function of the center-of-mass kinetic energy. Results are shown for models A, B and C. The band shows the extrapolation to the physical pion mass 
of the recent results of the HAL QCD  Collaboration \cite{Miyamoto:2017tjs} made by Haidenbauer and Krein in Ref.\ \cite{Haidenbauer:2017dua}.}
\label{fig:phs}
\end{center}
\end{figure}

\begin{figure}[t]
\begin{center}
\includegraphics[width=9.0cm]{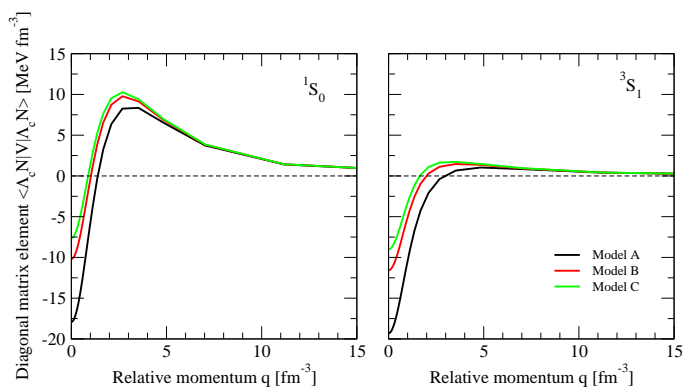}
\caption{(color on-line) $^1$S$_0$ and $^3$S$_1$  $\Lambda_cN\rightarrow \Lambda_cN$ diagonal matrix element as a function of the relative momentum q. Results are shown for models A, B and C. }
\label{fig:matel}
\end{center}
\end{figure}

The three $Y_c$N interaction models have then been used to solve the coupled-channel ($\Lambda_cN$, $\Sigma_cN$) Lipmann--Schwinger equation to obtain several scattering observables from the corresponding scattering amplitudes. The $\Lambda_cN$ phase shifts in the $^1$S$_0$ and $^3$S$_1$ partial waves are shown as a function of the center-of-mass kinetic energy in Fig.\ \ref{fig:phs} for the three models. The extrapolation to the physical pion mass of the recent results of the HAL QCD  Collaboration \cite{Miyamoto:2017tjs} made by Haidenbauer and Krein in Ref.\ \cite{Haidenbauer:2017dua} is shown by the green band for comparison. One can clearly see from the phase shifts that model A predicts a more attractive $\Lambda_cN$ interaction in the $^1$S$_0$ and $^3$S$_1$ partial waves  than the one derived in Ref.\ \cite{Haidenbauer:2017dua}. The reduction of the $g_{\Lambda_c\Lambda_c\sigma}$ and $g_{\Sigma_c\Sigma_c\sigma}$ couplings in models B and C leads to a reduction of attraction in these two partial waves which translates into a qualitatively better agreement between the phase shifts predicted by these two models and those obtained from the interaction of Ref.\ \cite{Haidenbauer:2017dua}, particularly in the low energy region.
\begin{table}[ht!]
\begin{center}
\begin{tabular}{c|cccc}
\hline
\hline
                      & Model A & Model B & Model C  & Ref.\ \cite{Haidenbauer:2017dua} \\ 
\hline
 $a_s$          & $-2.60$ & $-1.11$ & $-0.84$  & $-0.85 \cdot\cdot\cdot -1.00$\\
 $r_s$           & $2.86$ & $4.40$ & $5.38$ & $2.88 \cdot\cdot\cdot 2.61$ \\
                 $$ & $$           & $$ & $$   & $$ \\   
$a_t$          & $-15.87$ & $-1.52$ & $-0.99$ & $-0.81 \cdot\cdot\cdot -0.98$ \\
$r_t$           & $1.64$  & $2.79$ & $3.63$ & $3.50 \cdot\cdot\cdot 3.15$\\
 \hline
\hline

\end{tabular}
\end{center}
\caption{Singlet and triplet $\Lambda_cN$ scattering length and effective range predicted by the models A, B and C. The results of 
the extrapolation to the physical pion mass of the recent results of the HAL QCD  Collaboration \cite{Miyamoto:2017tjs} made by Haidenbauer and Krein in Ref.\ \cite{Haidenbauer:2017dua} are shown in the last column. Units are given in fm.}
\label{tab:tab2}
\end{table}
Note that the interaction derived in \cite{Haidenbauer:2017dua} predicts similar phase shifts for both partial waves since the corresponding $^1$S$_0$ and $^3 $S$_1$ potentials are almost identical, a feature already noted by the HAL QCD Collaboration at different values of the pion mass (see Ref.\ \cite{Miyamoto:2017tjs}) that seems to persist when extrapolating the lattice results to the physical point. 
This, however, is not the case of our models A, B and C which predict more overall attraction in the $^3$S$_1$ partial wave as it can be seen for example in Fig.\ \ref{fig:matel} where we show the diagonal $^1$S$_0$ and $^3 $S$_1$ matrix element in momentum space of the $\Lambda_cN\rightarrow \Lambda_cN$ channel. 

For completeness we report in Table \ref{tab:tab2} the singlet and triplet $\Lambda_cN$ scattering length and the effective range predicted by the three models. The results obtained by Haidenbauer and Krein in Ref.\ \cite{Haidenbauer:2017dua} are shown for comparison in the last column of the table. There is a good agreement between model C and the result of \cite{Haidenbauer:2017dua} for both scattering lengths. However,  is it pointed out in Ref.\ \cite{Haidenbauer:2017dua}  that the scattering lengths at the physical pion mass could in fact be as larger as $-1.3$ fm if the uncertainty of $\pm 0.2$ fm, given by the HAL QCD Collaboration for their result at $m_\pi=410$ MeV, is combined with the observation that variations in the scattering lengths of $\pm 0.05$ fm at this value of the pion mass amount to differences of about $\pm0.1$ fm at $m_\pi=138$ MeV. In this case, the prediction of model B would be in better agreement with the result of Haidenbauer and Krein than model C. 
Model A predicts a singlet effective range compatible with that obtained in Ref.\ \cite{Haidenbauer:2017dua} although a smaller triplet one. On the other hand, models B and C give a singlet effective range larger than that of \cite{Haidenbauer:2017dua}  but their agreement is qualitatively better for the triplet one.



\section{$\Lambda_c$ single-particle properties in finite nuclei}
\label{sec:sec3}

Here we briefly describe a method to obtain the  $\Lambda_c$ single-particle energies in a finite nucleus using 
an effective in-medium $Y_cN$ interaction derived from the bare $Y_cN$ potential presented in 
the previous section. The starting point of this method is the calculation of all the $Y_cN$ $G$-matrices, which describe the interaction between 
a charmed baryon ($Y_c=\Lambda_c,\Sigma_c$) and a nucleon in infinite nuclear matter. The $G$-matrices 
are obtained by solving the coupled-channel Bethe--Goldstone equation, written schematically as
\begin{eqnarray}
G_{Y_cN\rightarrow Y_c'N'}(\omega)=V_{Y_cN\rightarrow Y_c'N'}+\sum_{Y_c''N''}V_{Y_cN\rightarrow Y_c''N''} \nonumber \\
\times \frac{Q_{Y_c''N''}}{\omega-\epsilon_{Y_c''}-\epsilon_{N''}+i\eta}G_{Y_c''N''\rightarrow Y_c'N'}(\omega) \ ,
\label{eq:gfn}
\end{eqnarray}
where $V$ is the bare $Y_cN$ potential derived in
the previous section, $Q$ is the Pauli operator, that prevents the nucleon in the intermediate
state $Y_c''N''$ from being scattered below the Fermi momentum $k_{F_N}$, and 
$\omega$ is the nuclear matter starting energy which corresponds to the sum of the masses and
the  non-relativistic energies of the interacting charmed baryon and nucleon. 
We note that 
the Bethe--Goldstone equation has been solved in momentum space including partial waves up to a maximum value of the total angular momentum $J=4$.
We note also here that the so-called 
discontinuous prescription has been adopted, {\it i.e.,}
the  single-particle energies $\epsilon_{Y_c''}$ and $\epsilon_{N''}$ in the denominator of Eq.\ (\ref{eq:gfn}) are simply taken as the sum of the non-relativistic kinetic energy plus the mass of the corresponding baryon. 

The finite nucleus $Y_cN$ $G$-matrix, $G_{FN}$, can be obtained, in principle, by solving the Bethe--Goldstone equation directly in the finite nucleus \cite{hao93,halderson93}. The Bethe--Goldstone equation in finite nucleus is formally identical to Eq.\ (\ref{eq:gfn}), the only difference being the intermediate particle-particle propagator ({\it i.e.,} Pauli 
operator and energy denominator), which corresponds to that in the finite nucleus. Alternatively, one can find the appropiate $G_{FN}$ by relating it to the 
nuclear matter $Y_cN$ $G$-matrix already obtained. Eliminating the bare interaction $V$ in both finite nucleus and nuclear matter Bethe--Goldstone equations it is not difficult to write
$G_{FN}$ in terms of $G$ through the following integral equation:
\begin{eqnarray}
G_{FN}&=&G+G\left[\left(\frac{Q}{E}\right)_{FN}-\left(\frac{Q}{E}\right)\right]G_{FN} \nonumber \\
&=&G+G\left[\left(\frac{Q}{E}\right)_{FN}-\left(\frac{Q}{E}\right)\right]G \nonumber \\
&+&G\left[\left(\frac{Q}{E}\right)_{FN}-\left(\frac{Q}{E}\right)\right]G\left[\left(\frac{Q}{E}\right)_{FN}-\left(\frac{Q}{E}\right)\right]G \nonumber \\
&+& \cdot \cdot \cdot \ ,
\label{eq:gfng}
\end{eqnarray}
which involves the nuclear matter $G$-matrix and the difference between the finite nucleus and the 
nuclear matter propagators, written schematically as $(Q/E)_{FN}-(Q/E)$. This difference, which accounts for the relevant 
intermediate particle-particle states has been shown to be quite small (see Refs.\ \cite{borromeo92,morten94,morten96,vidana98,vidana00,vidana17})  and, therefore, 
in all practical calculations $G_{FN}$ can be well approximated by truncating the expansion (\ref{eq:gfng}) up second order in 
the nuclear matter $G$-matrix. Therefore, we have
\begin{equation}
G_{FN} \approx G+G\left[\left(\frac{Q}{E}\right)_{FN}-\left(\frac{Q}{E}\right)\right]G \ .
\label{eq:g2nd}
\end{equation}

\begin{figure}[t]
\begin{center}
\includegraphics[width=8.5cm]{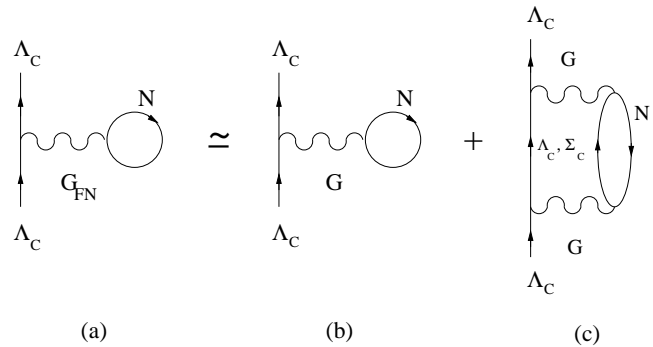}
\caption{Brueckner--Hartree--Fock approximation to the finite nucleus $\Lambda_c$ self-energy (diagram (a)), split into the sum of a
first-order contribution (diagram (b)) and a second-order 2p1h correction (diagram (c)).}
\label{fig:selfener}
\end{center}
\end{figure}

The finite nucleus $\Lambda_c$ self-energy can be obtained in the so-called Brueckner--Hartree--Fock approximation using the $G_{FN}$  as an effective 
$Y_cN$ interaction, as it is shown in diagram (a) of Fig.\ \ref{fig:selfener}. According to Eq.\ (\ref{eq:g2nd}) it can be split into the sum of the 
diagram (b), which represents the first-order term on the right-hand side of Eq.\ (\ref{eq:g2nd}), and the diagram (c), which stands for 
the so-called two-particle-one-hole (2p1h) correction, where the intermediate particle-particle propagator has to be viewed as the difference of propagators 
appearing in Eq.\ (\ref{eq:g2nd}). Schematically, it reads
\begin{eqnarray}
\Sigma_{BHF}=\sum_N \langle \Lambda_c N | G_{FN} | \Lambda_c N \rangle \approx \sum_N \langle \Lambda_c N | G | \Lambda_c N \rangle \,\,\,\,\,\,\, \nonumber \\
+\sum_{Y_cN} \langle \Lambda_c N | G | Y_cN \rangle \left[\left(\frac{Q}{E}\right)_{FN}-\left(\frac{Q}{E}\right)\right]
\langle Y_cN | G | \Lambda_c N \rangle \ . \nonumber \\
\label{eq:selfbhf}
\end{eqnarray}
Detailed expressions for the first-order and the 2p1h contributions to $\Sigma_{BHF}$ can be derived in close analogy to those for the finite 
nucleus $\Lambda$ self-energy given in Refs.\ \cite{morten96,vidana98,vidana00,vidana17} being, in fact,  formally identical. The interested reader
is referred to these works for details on the derivation and specific expressions of both contributions.

Finally, the self-energy can then be used as an effective $\Lambda_c$--nucleus mean field potential in a Schr\"{o}dinger equation in order to obtain the  energies and wave functions of the bound 
states of the $\Lambda_c$ in a finite nucleus. The Schr\"{o}dinger equation is solved by diagonalizing the corresponding single-particle
Hamiltonian in a complete basis within a spherical box following the procedure outlined in detail in Refs.\  \cite{borromeo92, morten94,morten96,vidana98,vidana00,vidana17}. 
Note that the Hamiltonian includes also the Coulomb potential since the $\Lambda_c$ is a positively charged baryon.

\section{Results}
\label{sec:sec4}

\begin{table*}[ht!]
\begin{center}
\begin{tabular}{c|ccc|c|ccc|c|ccc|c}
\hline
\hline
 
  & \multicolumn{3}{c}{$^{5}_{\Lambda_c}$He } & \multicolumn{1}{c}{$^{5}_{\Lambda}$He }  &  \multicolumn{3}{c}{$^{13}_{\Lambda_c}$C } & \multicolumn{1}{c}{$^{13}_{\Lambda}$C } & \multicolumn{3}{c}{$^{17}_{\Lambda_c}$O} & \multicolumn{1}{c}{$^{17}_{\Lambda}$O}    \cr

\hline
                       
$$ & $$  & $$    & $$    & $$ & $$ & $$ & $$ & $$  & $$ & $$ & $$  & $$ \\  

& Model A & Model B & Model C & J$\tilde A$ & Model A & Model B & Model C & J$\tilde A$ & Model A & Model B & Model C & J$\tilde A$   \\ 

$$ & $$  & $$    & $$    & $$ & $$ & $$ & $$ & $$  & $$ & $$ & $$  & $$ \\

 $1 s_{1/2}$ & $-13.58$ & $-3.24$ &  $-1.05$ & $-1.49$ &  $-27.26$ & $-10.20$ & $-5.47$ & $-7.84$ & $-31.76$ & $-12.47$  & $-6.96$ & $-10.04$ \\ 
 $1 p_{3/2}$ & $-1.74$ & $-$ & $-$ & $-$ & $-14.91$ & $-2.13$ & $-$ & $-$  & $-19.99$ & $-4.32$  & $-0.51$ & $-0.33$\\ 
 $1 p_{1/2}$ & $-0.39$ & $- $& $-$ & $-$ & $-13.42$ & $-1.03$ & $-$ & $-$  & $-18.79$ & $-3.22$  & $-$ & $-0.35$ \\ 
 
  $1 d_{5/2}$ & $-$ & $-$ &  $-$ & $-$ &  $-4.10$ & $-$ & $-$ & $-$ &  $-9.02$ & $-$  & $-$ & $-$ \\ 
  $1 d_{3/2}$ & $-$ & $-$ &  $-$ & $-$ & $-2.13$ & $-$ & $-$ & $-$  & $-6.96$ & $-$  & $-$  & $-$\\ 

$$ & $$  & $$    & $$    & $$ & $$ & $$ & $$ & $$  & $$ & $$ & $$  & $$ \\  
 
$2 s_{1/2}$ & $-$ & $-$ &  $-$ & $-$  &$-3.59$ & $-$ & $-$ & $-$ & $-7.13$ & $-$  & $-$  & $-$ \\ 
 
\hline

       & \multicolumn{3}{c}{$^{41}_{\Lambda_c}$Ca } & \multicolumn{1}{c}{$^{41}_{\Lambda}$Ca }  & \multicolumn{3}{c}{$^{91}_{\Lambda_c}$Zr } & \multicolumn{1}{c}{$^{91}_{\Lambda}$Zr }& \multicolumn{3}{c}{$^{209}_{\Lambda_c}$Pb} &  \multicolumn{1}{c}{$^{209}_{\Lambda}$Pb}  \cr

\hline
                       
$$ & $$  & $$    & $$    & $$ & $$ & $$ & $$ & $$  & $$ & $$ & $$  & $$ \\  

& Model A & Model B & Model C & J$\tilde A$ & Model A & Model B & Model C & J$\tilde A$ & Model A & Model B & Model C & J$\tilde A$   \\ 

$$ & $$  & $$    & $$    & $$ & $$ & $$ & $$ & $$  & $$ & $$ & $$  & $$ \\  

$1 s_{1/2}$ &  $-41.09$ & $-16.89$ & $-9.60$ & $-17.33$ & $-44.76$ & $-18.46$ & $-10.51$ & $-24.61$ & $-52.52$ & $-20.33$ & $-10.32$ & $-31.41$   \\ 
$1 p_{3/2}$ &  $-32.39$ & $-10.41$ & $-4.13$ & $-7.67$ & $-39.60$ & $-14.27$ & $-6.75$ & $-17.66$ &$-49.06$ & $-18.28$ & $-8.82$ & $-27.59$  \\ 
$1 p_{1/2}$ &  $-31.60$ & $-9.67$ & $-3.42$ & $-7.78 $ & $-39.24$ & $-14.00$ & $-6.49$ & $-17.58$ & $-48.84$ & $-18.10$ & $-8.64$ & $-27.58$   \\ 
$1 d_{5/2}$ &  $-23.10$ & $-3.91$ & $-$ & $- $& $-33.74$ & $-9.63$ & $-2.57$ & $-9.12$ & $-42.37$ & $-12.94$ & $-4.25$ & $-19.24$\\  
$1 d_{3/2}$ &  $-21.84$ & $-2.74$ & $-$ & $-$ & $-33.17$ & $-9.01$ & $-1.95$ & $-8.91$ & $-41.97$ & $-12.58$ & $-3.88$ & $-19.20$\\                  
$1 f_{7/2}$  &  $-13.54$ & $-$ & $-$ & $-$ & $-27.06$ & $-4.65$ & $-$ & $-1.35$ & $-37.47$ & $-9.11$ & $-0.59$ & $-10.51$ \\                   
$1 f_{5/2}$  & $-11.82$ & $-$  & $-$ &$-$ & $-26.29$ & $-3.80$ & $-$ & $-1.13$ & $-37.07$ & $-8.65$  & $-0.10$ &$-10.41$ \\           

$$ & $$  & $$    & $$    & $$ & $$ & $$ & $$ & $$  & $$ & $$ & $$  & $$ \\  

$2 s_{1/2}$  & $-20.47$ & $-2.74$  & $-$ & $-$ & $-31.13$ & $-8.05$ & $-1.29$ & $-6.60$ & $-40.53$ & $-10.20$  & $-1.13$ & $-17.43$ \\                    
$2 p_{3/2}$  & $-10.20$ & $-$ & $-$ & $-$ &$-22.81$ & $-2.23$ & $-$ & $-0.39$ &$-39.21$ & $-9.28$ & $-0.03$ & $-7.68$ \\ 
$2 p_{1/2}$  & $-9.24$ & $-$ & $-$ & $-$ &$-22.24$ & $-1.45$ & $-$ & $-0.38$ & $-38.95$ & $-9.06$ & $-$ & $-7.60$ \\   
$2 d_{5/2}$  & $-2.04$ & $-$ & $-$ & $-$ & $-14.62$ & $-$ & $-$ & $-$ & $-30.28$ & $-5.36$ & $-$ & $-4.85$ \\  
$2 d_{3/2}$  & $-0.95$ & $-$ & $-$ & $-$ &$-14.03$ & $-$ & $-$ & $-$ &$-29.83$ & $-4.75$ & $-$  & $-4.79$ \\
$2 f_{7/2}$   & $-$ & $-$ & $-$ & $-$ &$-7.90$ & $-$ & $-$ & $-$ & $-22.57$ & $-$ & $-$ &$-$ \\ 
$2 f_{5/2}$   & $-$ & $-$ & $-$ & $-$ &$-6.81$ & $-$ & $-$ & $-$ &$-22.10$ & $-$ & $-$ &$-$ \\  

$$ & $$  & $$    & $$    & $$ & $$ & $$ & $$ & $$  & $$ & $$ & $$  & $$ \\  

$3 s_{1/2}$  & $-1.15$ & $-$ & $-$ & $-$ & $-13.41$ & $-$ & $-$ & $-$ &$-23.80$ & $-1.51$ & $-$ & $-3.59$ \\  
$3 p_{3/2}$  & $-$ & $-$ & $-$ &$-$ & $-5.65$ & $-$ & $-$ & $-$ & $-22.32$ & $-$ & $-$  & $-$\\  
$3 p_{1/2}$   & $-$ & $-$ & $-$ & $-$ &$-5.61$ & $-$ & $-$ & $-$ &$-21.95$ & $-$ & $-$ & $-$ \\
$3 d_{5/2}$   & $-$ & $-$ & $-$ &  $-$ & $-$ & $-$ & $-$ & $-$& $-19.05$ & $-$ & $-$  & $-$\\
$3 d_{3/2}$   & $-$ & $-$ & $-$ & $-$  & $-$ & $-$ & $-$  & $-$ &$-18.33$ & $-$ & $-$ &  $-$ \\
$3 f_{7/2}$    & $-$ & $-$ & $-$ & $-$  & $-$ & $-$ & $-$ &  $-$ &$-5.58$ & $-$ & $-$  & $-$\\
 $3 f_{5/2}$   & $-$ & $-$ & $-$ & $-$. & $-$ & $-$ & $-$ &  $-$ &$-5.02$ & $-$ & $-$  & $-$ \\

$$ & $$  & $$    & $$    & $$ & $$ & $$ & $$ & $$  & $$ & $$ & $$  & $$ \\  

$4 s_{1/2}$  & $-$ & $-$ & $-$ & $-$& $-$ & $-$ & $-$ & $-$ & $-14.31$ & $-$ & $-$ & $-$\\  
$4 p_{3/2}$  & $-$ & $-$ & $-$ & $-$& $-$ & $-$ & $-$ & $-$ & $-1.19$ & $-$ & $-$ & $-$\\  
$4 p_{1/2}$  & $-$ & $-$ & $-$ & $-$& $-$ & $-$ & $-$ & $-$ & $-0.78$ & $-$ & $-$ & $-$\\  
$4 d_{5/2}$  & $-$ & $-$ & $-$ & $-$& $-$ & $-$ & $-$ & $-$ & $-0.68$ & $-$ & $-$ & $-$\\  

$$ & $$  & $$    & $$    & $$ & $$ & $$ & $$ & $$  & $$ & $$ & $$  & $$ \\  

$5 s_{1/2}$  & $-$ & $-$ & $-$ & $-$& $-$ & $-$ & $-$ & $-$ & $-0.52$ & $-$ & $-$ & $-$\\

\hline
\hline

\end{tabular}
\end{center}
\caption{Energy of $\Lambda_c$ single-particle bound states of several charm nuclei from $^5_{\Lambda_c}$He to $^{209}_{\Lambda_c}$Pb
obtained for the three models considered. Results for the single-particle bound states of the $\Lambda$ hyperon in the corresponding hypernuclei predicted by the original J\"{u}lich $\tilde A$ $YN$ interaction are also
shown for comparison. Units are given in MeV.}
\label{tab:bound}
\end{table*}

The energy of $\Lambda_c$ single-particle bound states in $^{5}_{\Lambda_c}$He, $^{13}_{\Lambda_c}$C , $^{17}_{\Lambda_c}$O, $^{41}_{\Lambda_c}$Ca, $^{91}_{\Lambda_c}$Zr  and $^{209}_{\Lambda_c}$Pb are shown in Table \ref{tab:bound} for the three models considered. For comparison the energy of the single-particle bound states of the $\Lambda$ hyperon in the corresponding hypernuclei, obtained with the original J\"{u}lich $\tilde A$ $YN$ interaction using the method described in the previous section, are also reported in the table. Note that all charmed nuclei (hypernuclei) considered consist of a closed shell nuclear core plus a $\Lambda_c$ ($\Lambda$) sitting in a single-particle state. Model A predicts the most attractive $\Lambda_cN$ interaction and, therefore, it predicts 
$\Lambda_c$ single-particle states more bound than models B and C, and a larger number of them as it can be seen in the table. Note that in the lack of experimental data on $\Lambda_c-$nuclei we cannot say {\it a priori} which one of the three models is better. However, since models B and C predict, as we saw before, scattering observables in better agreement with those extrapolated from LQCD in Ref.\ \cite{Haidenbauer:2017dua}, it would not be too risky to state that these two models are probably more realistic than model A.  


\begin{figure}[t]
\centering
\includegraphics[height=15.cm,angle=0]{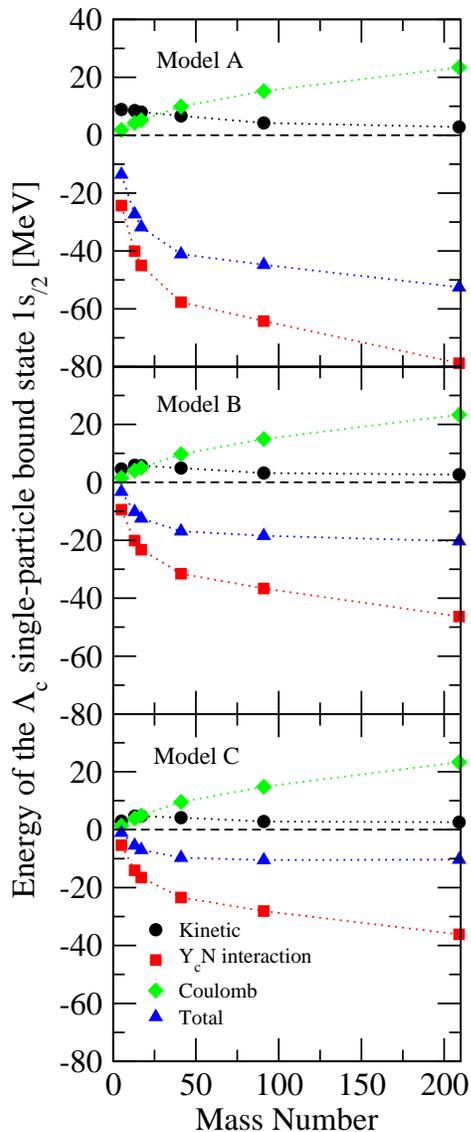}
\vspace{0.25cm}
\caption{(color on-line) Contributions of the kinetic energy,  the $Y_cN$ interaction and the Coulomb potential to the energy of the $\Lambda_c$ single-particle bound state $1s_{1/2}$ as a function of the mass number of the $\Lambda_c$ nuclei considered.}
\label{fig:expval}
\end{figure}


\begin{table}[ht!]
\begin{center}
\begin{tabular}{c|ccc|c}
\hline
\hline

& Model A & Model B & Model C & J$\tilde A$    \\ 

\hline
                      
$1 s_{1/2}$ & $-31.54$  & $-12.57$    & $-7.11$    & $-8.78$ \\ 
$1 p_{3/2}$ & $-19.69$  & $-4.37$    & $-0.58$    & $-$\\ 
$1 p_{1/2}$ & $-18.45$  & $-3.24$    & $-$    & $-$ \\ 
$1 d_{5/2}$ &  $-8.71$  & $-$    & $-$    & $-$\\ 
$1 d_{3/2}$ & $-6.62$  & $-$    & $-$    & $-$ \\ 

$$ & $$  & $$    & $$    & $$  \\
 
$2 s_{1/2}$ & $-7.02$  & $-$    & $-$    & $-$ \\ 
 
\hline
\hline

\end{tabular}
\end{center}
\caption{Energy of $\Lambda_c$ single-particle bound states of  $^{17}_{\Lambda_c}$O when the coupling of the $\Lambda_cN$ and the $\Sigma_cN$ channels is switched off. Results for the $\Lambda$ hyperon in $^{17}_{\Lambda}$O obtained with the original J\"{u}lich $\tilde A$ $YN$ interaction are also shown for comparison. Units are given in MeV.}
\label{tab:scn}
\end{table}


Looking now back at the table we observe (as in the case of single $\Lambda-$hypernuclei) a small spin-orbit splitting of the $p-,d-$ and $f-$wave states in all $\Lambda_c-$nuclei, specially in the case of the heavier ones where it is of the order of few tenths of MeV.  In addition, we also note that, since the $\Lambda_c$ is heavier than the $\Lambda$, the level spacing of the $\Lambda_c$ single-particle energies is, for the three models, always smaller than the corresponding one for the hypernuclei.  These observations are in agreement with the results previously obtained by in Tsushima and Khanna in Refs.\ \cite{tsushima03b,tsushima03c,tsushima03d} using the quark-meson coupling model and, later, by Tan and Ning in Ref.\ \cite{tan04} within a relativistic mean field approach. Although there exist formal differences between our calculation and those of Refs.\ \cite{tsushima03b,tsushima03c,tsushima03d,tan04} that give rise to different predictions for the $\Lambda_c$ single-particle bound states in finite nuclei, our results (particularly those for models B and C) are in general compatible with those of these works (see {\it e.g.,} tables I and II of Ref.\ \cite{tsushima03b} and table I of Ref.\ \cite{tan04}). 


\begin{figure*}[t]
\centering
\includegraphics[height=8.cm,angle=0]{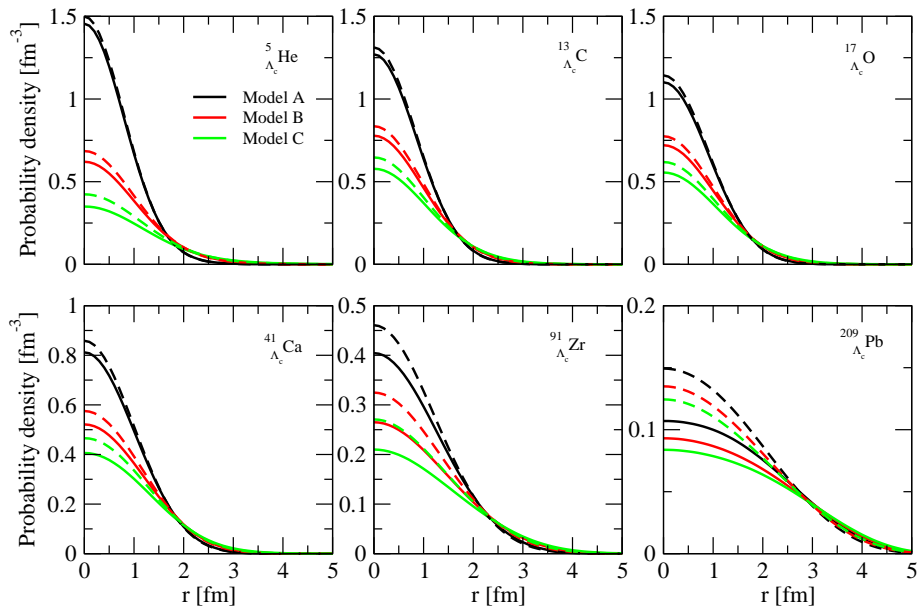}
\vspace{0.25cm}
\caption{(color on-line) $\Lambda_c$ probability density distribution for the $1s_{1/2}$ state in the six $\Lambda_c-$nuclei considered. Results are shown for the three models A, B and C of the $Y_cN$ interaction. Dashed lines show the result when the Coulomb interaction is artificially switched off.}
\label{fig:wf}
\end{figure*}


It has been pointed in Refs.\ \cite{tsushima03b,tsushima03c,tsushima03d,tan04} and, more recently,  also in Ref.\ \cite{Miyamoto:2017tjs} that the Coulomb interaction plays a non-negligible role in $\Lambda_c-$nuclei, and that their existence is only possible if their binding energy is larger than the Coulomb repulsion between the $\Lambda_c$ and the protons. To understand better the effect of the Coulomb force in our calculation, in Fig.\ \ref{fig:expval} we explicitly show the contributions of 
the kinetic energy,  the $Y_cN$ interaction and the Coulomb potential to the energy of the $\Lambda_c$ single-particle bound state $1s_{1/2}$ as a function of the mass number ($A=N+Z+1$, with $N$ the neutron number and $Z$ the atomic number) of the $
\Lambda_c-$nuclei considered. Note that, while the Coulomb contribution increases because of the increase of the number of protons with the atomic number, those of the kinetic energy and the $Y_cN$ interaction decrease when going from light to heavy $\Lambda_c-$nuclei. The kinetic energy contribution decreases with the mass number because the wave function of the $1s_{1/2}$ state (see Fig.\ \ref{fig:wf}) becomes more and more spread due to the larger extension of the nuclear density over which the $\Lambda_c$ wants to be distributed. The increase of the nuclear density with $A$ leads to a more attractive $\Lambda_c$ self-energy (see {\it e.g.,} figures 2 and 3 of Ref.\ \cite{vidana17} for a detail discussion in the case of  single $\Lambda-$hypernuclei) that translates into a more negative contribution of the $Y_cN$ interaction. Note that, when adding the three contributions, the energy of the $1s_{1/2}$ bound state decreases by several MeV in the low mass number region and then it tends to saturate (being almost constant for model C) for heavier nuclei. This is due to a compensation  between the attraction of the $Y_cN$ interaction and the repulsion of the Coulomb force. We note that this compensation leads, particularly in the case of model B, to values of the $\Lambda_c$ single-particle bound state energies similar to those obtained for  single $\Lambda$-hypernuclei with the J\"{u}lich $\tilde A$ $YN$ potential (see Table \ref{tab:bound}). We want to point out that even the less attractive one of our $Y_cN$ interactions (model C), despite the Coulomb repulsion, is able to bind the $\Lambda_c$ in all the nuclei considered.  This is in contrast with the recent results of the HAL QCD Collaboration \cite{Miyamoto:2017tjs} which suggest that only light or medium-mass $\Lambda_c-$nuclei could really exist. However, we note that this conclusion is based on results obtained for a value of the pion mass of 410 MeV which give rise to a $Y_cN$ interaction much less attractive than ours and the one derived in Ref.\ \cite{Haidenbauer:2017dua} when these lattice results are extrapolated to the physical pion mass (see figures 1 and 2 of Ref.\ \cite{Haidenbauer:2017dua}).

Now we would like to focus the attention of the reader for a while on the effect of the coupling of the $\Lambda_c N$ and $\Sigma_c N$ channels. These two channels are located at approximately $3224$ and $3394$ MeV, respectively. Being separated by about $170$ MeV it is expected, as it was already pointed out by Tsushima and Khanna (see {\it e.g.,} \cite{tsushima03b,tsushima03c}), that the effect of their coupling on charmed nuclei will be less important than that  of the $\Lambda N$ and $\Sigma N$ channels (separated only by $\sim 80$ MeV) on hypernuclei. This is illustrated in Table \ref{tab:scn} where we show as an example the energy of the $\Lambda_c$ ($\Lambda$) single-particle states bound states of  $^{17}_{\Lambda_c}$O ($^{17}_{\Lambda}$O) when the $\Lambda_cN-\Sigma_cN$ ($\Lambda N-\Sigma N$) coupling is switched off.  Note that  the differences between the levels obtained with the complete coupled-channel calculation for  $^{17}_{\Lambda_c}$O (see Table \ref{tab:bound}) and without the $\Lambda_c N-\Sigma_c N$ coupling are almost negligible, being of the order of few tenths of MeV or less, whereas those for $^{17}_{\Lambda}$O are slightly larger than $1$ MeV. Note also that the elimination of the coupling between the $\Lambda_c N$ and $\Sigma_c N$ channels leads, in the case of models B and C, to a bit more of attraction, contrary to what happens in the hypernuclei case where the $\Lambda$ bound states become less bound when the $\Lambda N-\Sigma N$ coupling is eliminated. 

We finish this section by showing in Fig.\ \ref{fig:wf}, for the three models, the probability density distribution ({\it i.e.,} the square of the radial wave function) of the $\Lambda_c$ in the $1s_{1/2}$ state for the six $\Lambda_c-$nuclei considered. The result when the Coulomb interaction is artificially schitwed off is also shown for comparison (dashed lines). Note that, due to the increase of the nuclear density, when moving from light to heavy nuclei the probability density of finding the $\Lambda_c$ close to the center of the nucleus decreases, and it becomes more and more distributed over the whole nucleus. Note also that, as expected, due to the Coulomb repulsion the $\Lambda_c$ is pushed away from the center of the nuclei. A similar discussion can be done for the probability densities  of the the other $\Lambda_c$ single-particle bound states.


\section{Summary and conclusions}
\label{sec:sec5}

In this work we have determined the single-particle energies of the $\Lambda_c$ charmed baryon in several nuclei. To such end, we have developed a charmed baryon-nucleon interaction based on a SU(4) extension of the meson-exchange hyperon-nucleon potential 
$\tilde A$ of the J\"{u}lich group. We have considered three different models (A, B and C) of this interaction that differ only on the values of the couplings of the scalar $\sigma$ meson with the charmed baryons. Several scattering observables have been computed with the three models and compared with those predicted by the $Y_cN$ interaction derived by Haidenbauer and Krein \cite{Haidenbauer:2017dua} from the extrapolation to the physical pion mass of the recent results of the HAL QCD Collaboration \cite{Miyamoto:2017tjs}.  Qualitative agreement has been found between the predictions of our models B and C and those of the model by Haidenbauer and Krein \cite{Haidenbauer:2017dua}. 

The three models have then been used to obtain the self-energy of the $\Lambda_c$ in finite nuclei by using a many-body approach that started with the construction of a nuclear matter $Y_cN$ $G$-matrix from which a finite nucleus one was derived through a perturbative expansion. Using the resulting $\Lambda_c$ self-energy as an effective $\Lambda_c-$nucleus mean field potential in a Schr\"{o}dinger equation we have finally obtained the energies and wave functions of the bound states of the $\Lambda_c$ in the different nuclei.

Our results (particularly those for models B and C) are compatible with those obtained by Tshushima and Khanna \cite{tsushima03b,tsushima03c,tsushima03d} and Tan and Ning \cite{tan04}, despite the formal differences between our calculation and those of these works based, respectively, on the quark-meson coupling model and the relativistic mean field approach. A small spin-orbit splitting of the $p-, d-$ and $f-$wave states has been found as in the case of single $\Lambda$-hypernuclei. It has been also observed that level spacing of the $\Lambda_c$ single-particle energies is smaller than the corresponding one for hypernuclei. 

We have analyzed the role played by the Coulomb potential  on the energies of the $\Lambda_c$ single-particle bound states. This analysis has shown that the compensation between the $Y_cN$ interaction and the repulsion of the Coulomb force leads, particularly in the case of model B, to values of the $\Lambda_c$ single-particle bound state energies similar to those obtained for the single $\Lambda$-hypernuclei with the original J\"{u}lich $\tilde A$ $YN$ potential. The analysis has also shown that, despite the Coulomb repulsion, even the less attractive one of our $Y_cN$ interactions (model C) is able to bind the $\Lambda_c$ in all the nuclei considered. This is in contrast with the recent results of the HAL QCD Collaboration \cite{Miyamoto:2017tjs} which suggest that only light or medium-mass $\Lambda_c-$nuclei could really exists. However,  the conclusion of this work is based on results obtained for a value of the pion mass of 410 MeV which give rise to a $Y_cN$ interaction much less attractive than ours and the one derived in Ref.\ \cite{Haidenbauer:2017dua} when these lattice results are extrapolated to the physical pion mass.

Finally, we have shown that the effect of the coupling of the $\Lambda_cN$ and $\Sigma_cN$ channels on the single-particle properties of charmed nuclei is much less important (being in fact almost negligible) than that of the $\Lambda N$ and $\Sigma N$ channels on the corresponding properties of single $\Lambda$-hypernuclei, due to the large mass difference of the $\Lambda_c$ and $\Sigma_c$ baryons of $\sim 170$ MeV.


\section*{Acknowledgments}

The authors are very grateful to Johann Haidenbauer for his useful comments. This work has been partly supported by the COST Action CA16214 and by the Spanish Ministerio de Economia y Competitividad (MINECO) under the project MDM-2014-0369 of ICCUB (Unidad de Excelencia ``Mar\'\i a de Maeztu"), 
and, with additional European FEDER funds, under the project
FIS2017-87534-P.




\end{document}